\documentclass[twoside]{article}
\usepackage{fleqn,espcrc2}
\usepackage{graphicx}
\usepackage[figuresright]{rotating}

\newcommand{\AmS}{{\protect\the\textfont2
  A\kern-.1667em\lower.5ex\hbox{M}\kern-.125emS}}
\title{Non-Oscillation Searches of Neutrino Mass 
in the Age of Oscillations}
\author{Francesco Vissani\address{INFN, Laboratori Nazionali del Gran Sasso,
Theory Group, I-67010 Assergi (AQ), 
Italy}\thanks{I thank Christian Weinheimer, who 
proposed the task and helped me with several discussions.}}
\begin{document}
\begin{abstract}
We focus on the implications of the oscillations 
for the shape of nuclear $\beta$-spectrum
(=direct search for $\nu$ mass).
This is of interest because of the existing bound, 
$m_{\rm \nu_e}<2.2$ eV, that could improve by 
one order of magnitude with future experiments.
We stress important connections with the results of
Liquid Scintillator Neutrino Detector (LSND),
$\nu_{\rm e}$ disappearance experiments,
supernova (SN) neutrinos and neutrinoless 
double beta decay ($0\nu2\beta$).
\vspace{-1pc}
\end{abstract}
\maketitle
\section{Massive Neutrinos and $\beta$-Spectrum}
The nuclear $\beta$-spectrum with emission of $N$ 
(undetected and not too heavy) massive $\nu$'s is the 
weighted sum of individual 
$\beta$-spectra \cite{80,csz}:
$$
d\Gamma_{  mass}=\sum_{j=1}^{N}
|U_{{\rm e}j}^2|\times d\Gamma(m_j^2),\ \ \ m_j\le m_{j+1}
$$
What is the ``effective'' mass
$m_{\rm \nu_e}^2,$ for which  
$d\Gamma_{  mass}\sim d\Gamma(m_{\rm \nu_e}^2)$? 
Let us work out an answer. 
For $E_\nu\ll Q,$ the behaviour of the $\beta$-spectrum 
is just due to phase space:
$d\Gamma(m_j^2)\propto E_\nu (E_\nu^2-m_j^2)^{1/2} d E_\nu.$
If also the condition $E_\nu\gg m_j$ holds, we can
approximate $d\Gamma(m_j^2)\propto [E_\nu^2-m_j^2/2] d E_\nu.$ 
In these hypotheses, 
$d\Gamma_{  mass}\propto E_\nu^2 \sum_j |U_{{\rm e}j}^2|-
\sum_j |U_{{\rm e}j}^2| m_j^2/2] d E_\nu,$ so one is lead to 
define:
$$
m_{\rm \nu_e}^2\equiv \frac{\sum_{j} |U_{{\rm e}j}^2|
\times m_j^2}{\sum_{j} |U_{{\rm e}j}^2|}\ \ \  
({\mbox{or }} \equiv \sum_j |U_{{\rm e}j}^2|\times m_j^2)
$$
The simpler formula \cite{mai99}
(that we keep for reference) holds 
if unitarity is assumed;
or, in practice, if the normalization 
is not checked experimentally.
A warning: In endpoint type experiments, 
the sums on $j$ extend {\em just} on 
neutrino masses within the region of measurement
(typically much narrower than the $E_\nu$ 
range, $E_\nu\le Q$).

\begin{figure}[htb]
\includegraphics*[width=14pc,angle=270]{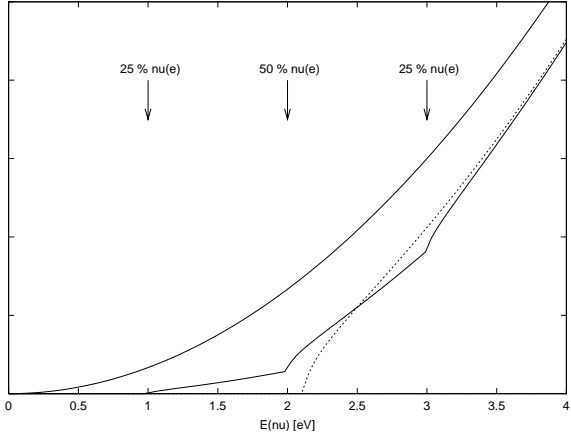}
\vskip-1pc
\caption{{\em Illustrative} endpoint spectrum,
shown as function of $E_\nu,$ in arbitrary 
unities. Arrows show the position 
of the mass levels, with indicated values of $|U_{{\rm e}j}^2|.$
True spectrum (continuous line) is compared with the 
spectrum with effective mass $m_{\rm \nu_e}^2$ (dotted). 
Massless spectrum is also shown for comparison. 
[A comment on two more {\em realistic} cases: 
1) for a {\tt SAL} 4-$\nu$ 
mass spectrum, an ``LSND peak'' at $\sim 1$ eV 
can be visually appreciated 
only after magnifying by some 100$\times$ 
on $E_\nu<\mbox{few eV}$; 
2) for a {\tt SA} 3-$\nu$ mass spectrum, 
to see ``ATM or SOLAR peaks'' requires 
further 1000$\times$ zooming on $E_\nu<100$ meV.]}
\label{fig1}
\vskip-1pc
\end{figure}

Is is likely to measure something
more than $m_{\rm \nu_e}^2?$ The answer is conditional;
{\em No}, if the energy resolution $\delta E_\nu$
is larger than the level spacing; for this would 
mean averaging 
the $\beta$-spectrum $\langle d\Gamma_{ mass} \rangle,$ 
reducing it to $\langle d\Gamma(m_{\rm \nu_e}^2)\rangle$ 
(far from endpoint the effective spectrum reproduces 
well the true one, as seen in fig.\ \ref{fig1}). 
{\em Maybe yes}, if one level (at least) stands out 
from the other ones. 
So the answer depends essentially on the 
detector characteristics.
To be specific, if we assume that 
$\nu_H$ has mass in the $5-10$ eV region, $m_{\rm \nu_e}^2$ 
could be not adequate to describe 
existing tritium endpoint data \cite{mai99,tro99}
(unless the mixing of $\nu_H$ with $\nu_{\rm e}$ is 
so little, that its effect is invisible). 
We assume that this does not happen; thus, a 
\underline{single parameter} describes the modifications of the 
nuclear $\beta$-spectrum.
However, we stress again 
that a single parameter suffices
{\em if} the spectrum is ``not resolved''.

The parameter $m_{\rm \nu_e}^2$ can be compared with 
${\cal M}_{\rm ee}^2=|\sum_j |U_{{\rm e}j}^2|\times m_j\times 
\exp(i\xi_j)|^2$ (the 
ee-entry of the Majorana neutrino mass matrix, squared)
that leads to $0\nu2\beta$ \cite{0nu2beta}. 
Note that:\newline
$\bullet$ The presence of the Majorana phases $\xi_j,$ that 
can produce cancellations in ${\cal M}_{\rm ee};$\newline
$\bullet$ The different dependence on $|U_{{\rm e}j}^2|:$
Individual contributions scale as 
$\delta {\cal M}_{\rm ee}=|U_{{\rm e}j}^2| \times m_j,$ while
$\delta m_{\rm \nu_e}=|U_{{\rm e}j}| \times m_j.$ Thence, the 
$1^{st}$ parameter is more severely suppressed 
than the $2^{nd}$ by $|U_{{\rm e}j}^2|.$ 

\section{The Connection with Oscillations}
Mainz \cite{mai99} and Troitsk \cite{tro99}
Collaborations pushed the limit on
$m_{\rm \nu_e}$ down to 2.2 eV, and there are plans under discussion
to reach the $200-400$ meV level. 
Being at the Neutrino Oscillation Workshop, 
it is natural to ask what we expect for 
$m_{\rm \nu_e}^2,$ if neutrino do oscillate. 
With little algebra:
\begin{equation}
\begin{array}{l}
m_{\rm \nu_e}^2=\sum_j |U_{{\rm e}j}^2|\times m_j^2=\\[1ex]
\ \ \sum_{j>1} |U_{{\rm e}j}^2|\times \Delta m_{j1}^2+m_1^2\equiv
\delta m^2 + m_1^2
\end{array}
\label{eq1}
\end{equation}
we separate the part of $m_{\rm \nu_e}^2$
that we can obtain from oscillations
(namely, $\delta m^2,$ which is $\ge 0$)
from the rest, irrelevant to oscillations
(namely, $m_1^2,$ the squared mass of the lightest neutrino).

To proceed, we will consider certain scenarios of oscillations
and \underline{neutrino mass spectra}. 
Motivated by existing indications of oscillation
\cite{now}, we select 6 cases with 4-$\nu$
($(N-1)$! level permutations)
and  6 cases with 3-$\nu$ (implies 
to discard one of the indications).
The caption of Tab.\ 1 and two examples 
should suffice to clarify our terminology:
1) The spectrum {\tt SAL} is the 4-$\nu$ spectrum, where 
the splitting $\Delta m^2_{21}$ is related to solar neutrinos,
$\Delta m^2_{32}$ to atmospheric neutrinos, $\Delta m^2_{43}$ to LSND.
2) The spectrum {\tt SA} does not account for LSND, and entails 
3-$\nu$. 
We assume that a ``sterile'' neutrino 
plays a role, but only in the 4-$\nu$ cases.
There are 3 main cases:
                       \newline{\em 1.} {$\delta m^2$ \sc large.}
This happens for the 4-$\nu$ spectra {\tt ALS}, 
{\tt LSA}, {\tt LAS} \cite{bil96},
and for the 3-$\nu$ spectra
{\tt AL} and {\tt LS}, where 
$\delta m=400-1400$ meV.
Indeed, the $\nu_{\rm e}$ state has to 
stand above the ``LSND mass gap'', because 
it must be involved in the solar doublet of levels,
or (only for the {\tt AL} 
\setcounter{footnote}{0}
case\footnote{This case is special, also
because it does not permit to arrange cancellations
for ${\cal M}_{\rm ee}.$ 
In the other four cases, it is instead {\em possible}
to arrange such a cancellation for $0\nu2\beta;$ 
but the solar mixing has to be large.})
because we know that
the atmospheric doublet is mostly $\nu_\mu-\nu_\tau.$
This is the most appealing case for the future experiments 
that aim at finding an effect of  massive neutrinos in $\beta$-decay
spectra. However, all these 5 spectra 
might have troubles
with SN1987A $\nu$'s \cite{sn87}:
In fact, the mixing of $\vartheta_{\rm e\mu}$
that we need to explain LSND 
produces resonant MSW 
conversion\footnote{The electron density in 
the SN core is so large 
({\em e.g.}, when compared with solar densities) 
that can lead to an MSW resonance  for 
any of the $\Delta m^2$'s, LSND and atmospheric 
included.} \cite{msw}
of $\overline{\nu}_{\mu }$ into $\overline{\nu}_{\rm e}.$
This implies an average energy of the dominant class of events
($\overline{\nu}_{\rm e} p\to {\rm e}^+ n$) significantly
larger than expected, that does not seem to 
be what data suggest. 
                             \newline{\em 2.} {$\delta m^2$ \sc medium.}
There are 5 sub-cases:
(1) The first two spectra are {\tt SAL} and {\tt ASL}.
For them, $\delta m\sim (\Delta m^2_{lsnd})^{1/2}\times \vartheta_{\rm ee}=
50-180$ meV. In fact, the mixing that lead to 
appearance in LSND is in these schemes \cite{bil96}
$\vartheta_{\rm e\mu}\approx \vartheta_{\rm ee}\times \vartheta_{\rm \mu\mu},$
the product of those mixings that 
would lead to disappearance 
in Bugey \cite{bugey}, $\vartheta_{\rm ee},$ and 
in CDHS \cite{cdhs}, $\vartheta_{\mu\mu}.$
(The {\em final} LSND data do not
contradict this, but the bound is almost saturated).
(2) Then we have the spectrum {\tt LA}. 
$\delta m$ is tunable up to $120-180$ meV, since 
one can arrange the lighter state to be
$\nu_1\approx \nu_{\rm e} + \vartheta_{\rm ee} \nu_\tau
+\vartheta_{\rm e\mu} \nu_\mu;$ the mixing
$\vartheta_{\rm e\mu}$ is fixed by LSND, while
$\vartheta_{\rm ee}$ (that is what matters for us)
is only loosely constrained by Bugey. 
(3) Next case is {\tt SL}; again, $\delta m$ is tunable.
If $\nu_3=\nu_\mu+\vartheta_{\rm e\mu}\nu_{\rm e} + ...,$
$\delta m=20-50$ meV, if 
$\nu_3=\nu_\tau+\vartheta_{\rm ee}\nu_{\rm e} + ...,$ instead,
$\delta m=120-180$ meV.
(4) The {\em naive} expectation for the {\tt SLA} spectrum is
$\delta m^2=\Delta m^2_{lsnd}\times \vartheta_{\rm e\mu}^2,$ but 
in fact $\delta m$ can be larger, if:
$$
\nu_{\rm e}\approx n+\vartheta_{\rm e\mu} N + 
\vartheta_{\rm ee} N_{\!\perp}, \ \
\nu_{\mu}\approx N+\vartheta_{\mu\mu} n_{\!\perp} - \vartheta_{\rm e\mu} n
$$
this case shows that
$\delta m^2$ can go up $\Delta m^2_{lsnd}\times \vartheta_{\rm ee}^2$
[$n$ is a linear combination of $\nu_1$ and $\nu_2,$
$N$ is a linear combination of $\nu_3$ and $\nu_4,$ and the 
orthogonal states have obvious notations], 
still keeping the probability of appearance 
$P_{\rm e\mu}\sim 4\times \vartheta_{\rm e\mu}^2\times \sin^2\varphi,$
and the probability of disappearance
$P_{\rm ee}\sim 1-4\times \vartheta_{\rm ee}^2\times \sin^2\varphi$
(5) Last case is the spectrum {\tt AS}, where 
the trivial identification $\delta m=(\Delta m^2_{atm})^{1/2}=40-80$ meV
holds. We remind the reader that this case is among the targets of 
next generation $0\nu2\beta$ experiments, indeed 
${\cal M}^2_{\rm ee}\ge \Delta m^2_{atm}\times 
(1-\sin^2 2\theta_{sol}).$
SN1987A bounds can be avoided
(at the price of a fine tuning of the relevant mixing).
                             \newline{\em 3.} {$\delta m^2$ \sc small.}
This includes only the 3-$\nu$ spectrum {\tt SA},
when $\delta m^2=\Delta m^2_{sol}\times |U_{\rm e2}^2| +
\Delta m^2_{atm}\times |U_{\rm e3}^2|=(2.5-20\mbox{ meV})^2$
(LMA has been assumed).  If solar oscillations will be confirmed,
but MiniBooNE should not support LSND findings, 
this case would be quite (most?) likely.
For easy reference, 
the results of the discussion above 
are reported in one table:
\begin{table}[hbt]
\vskip-1pc
\caption{Summary of the expectations on $\delta m$
(=lower bound on $m_{\rm \nu_e}$ from oscillations,
Eq.\ \ref{eq1}). 
The letters of the acronyms stand for 
{\tt L}=$\Delta m^2_{lsnd},$ 
{\tt A}=$\Delta m^2_{atm},$ {\tt S}=$\Delta m^2_{sol};$
the order in which they are written
(from left to right) indicates how the 
$\Delta m^2$'s appear in the given 
neutrino mass spectrum (from lighter to heavier one).\label{tab1}} 
\vskip.2cm
\begin{center}
\begin{tabular}{cc}
\hline
$\delta m$ [meV] & {\sc Spectrum}  \\
\hline
$400-1400$ & {\tt ALS,LSA,LAS,AL,LS} \\
$50-180$ & {\tt SAL,ASL} \\ 
$20-180$ & {\tt SLA,LA,SL} \\
$40-80$ & {\tt AS}   \\
$2.5-20$ & {\tt SA} \\
\hline
\end{tabular}
\end{center}
\vskip-12mm
\end{table}
\section{Discussion}
Oscillations lead to 
consider neutrino mass;
this could be related with 
$\mu\to {\rm e}\gamma,$
proton decay, {\em etc.} But perhaps, the {\em most direct}
connections are those with the $0\nu2\beta$ decay, and 
possible distortions of the $\beta$-decay spectra. In this view,
we considered the parameter $m_{\rm \nu_e}^2$ and discussed 
the expectations on that part of it, $\delta m^2,$ 
related with oscillations (Eq.\ \ref{eq1}).
We have outlined a troublesome connections of 
those schemes that predict the largest values 
of $\delta m^2$ with SN1987A neutrinos.
Still, $\delta m^2$ could be relatively large (perhaps
observable in future setups) if LSND indications
are due to $\bar{\nu}_{\rm e}$ appearance. 
This is strictly  connected with $\bar{\nu}_{\rm e}$ 
{\em disappearance}, since 
the implied mixing $\vartheta_{\rm ee}$ is the 
parameter that matters for $\beta$-decay spectra. 
If LSND signal is not due to oscillations, 
$m_{\rm \nu_e}^2$ and $m_1^2$ 
could be identified for practical purposes:
Only ``quasi degenerate'' neutrinos could significantly 
modify $\beta$-spectra, unless the 50 meV level 
is attained (which at present seems quite difficult). 

In conclusion, we stress again that the 
expectations for  $m_{\rm \nu_e}^2$ are 
closely related with oscillations: LSND indications 
of flavor appearance; neutrinos from SN1987A and 
future type I$\!$I SN's;
existence of sub-dominant $\nu_{\rm e}$ mixing...\
and, in many (but not all) cases, 
also with the rate of the neutrinoless double beta 
transition.

\end{document}